\documentclass{emulateapj}
\usepackage{hyperref,amsmath,amssymb,times,graphicx,xcolor,fleqn,enumitem}

\begin{document}

\title{Tidal disruption rates in non-spherical galactic nuclei formed by galaxy mergers}
\author{Kirill Lezhnin\altaffilmark{1}}
\email{klezhnin@yandex.ru, eugvas@lpi.ru}
\author{Eugene Vasiliev\altaffilmark{2,3}}
\affil{$^{1}$Moscow Institute of Physics and Technology, Institutskiy per. 9, Dolgoprudny, Moscow Region, Russia, 141700}
\affil{$^{2}$Rudolf Peierls Centre for Theoretical Physics, 1 Keble road, Oxford, UK, OX1 3NP}
\affil{$^{3}$Lebedev Physical Institute, Leninsky prospekt 53, Moscow, Russia, 119991.}

\newcommand{\jlc}{j_\mathrm{lc}}
\newcommand{\Jgap}{J_\mathrm{gap}}
\newcommand{\Jcirc}{J_\mathrm{circ}}
\newcommand{\ah}{a_\mathrm{h}}
\newcommand{\rinfl}{r_\mathrm{infl}}
\newcommand{\Trel}{t_\mathrm{rel}}
\newcommand{\Trefill}{t_\mathrm{refill}}

\begin{abstract}
We explore the rates of tidal disruption events (TDEs) of stars by supermassive black holes (SBHs) in galactic nuclei formed in mergers followed by a formation and coalescence of a binary SBH. Such systems initially have a deficit of stars on low-angular-momentum orbits caused by the slingshot process during the binary SBH stage, which tends to reduce the flux of stars into the SBH compared to the steady-state value. On the other hand, a newly formed galactic nucleus has a non-spherical shape which enhances the mixing of stars in angular momentum and thus the TDE rate. 
In galaxies with relatively low SBH masses ($\lesssim 10^7\,M_\odot$), relaxation times are short enough to wash out the anisotropy in initial conditions, and for more massive SBH the enhancement of flux due to non-sphericity turns out to be more important than its suppression due to initial anisotropy. Therefore, the present-day TDE rates generally exceed conventional steady-state estimates based on a spherical isotropic approximation. We thus conjecture that the lower observationally inferred TDE rates compared to theoretical predictions cannot be attributed to the depletion of low-angular-momentum stars by SBH binaries.
\end{abstract}

\section{Introduction}

Tidal disruption events (TDEs) are luminous short-living flares that are believed to happen in galactic nuclei containing supermassive black holes (SBHs): a close encounter between a star and SBH leads to the disruption of the star if the gradient of gravitational force is large enough \citep{Rees1988}. These events leave imprints in the entire electromagnetic spectrum: optical \citep{vanVelzenFarrar2014}, UV \citep{Gezari2008} or X-ray \citep{Donley2002, KhabibullinSazonov2014} observations imply an average rate of $10^{-4}-10^{-5}$ events per year per galaxy \citep[e.g.,][]{Komossa2015}. 
The loss-cone theory describing the flux of stars into a SBH was first developed in 1970s in the context of globular clusters \citep[e.g.,][]{FrankRees1976, LightmanShapiro1977}, and later applied to galactic nuclei \citep{SyerUlmer1999,MagorrianTremaine1999,WangMerritt2004,MageshwaranMangalam2015,Kochanek2016,Aharon2016}. Even though the TDE rates inferred from observations are of the same order as the theoretical estimates, the latter are systematically higher, as stressed by \citet{StoneMetzger2016},\defcitealias{StoneMetzger2016}{SM2016} hereafter SM2016. However, uncertainties are high on both sides, and the rate of \textit{observable} events depends both on their \textit{intrinsic} rate and the details of emission mechanisms in the process of tidal disruption itself. In this paper we deal only with the first factor, namely the stellar-dynamical estimate of the flux of stars into the loss cone of a SBH.

The simplest and most commonly used estimate is based on the steady-state solution of the Fokker--Planck equation describing the diffusion of stars in angular momentum for a spherically-symmetric stellar system, driven by two-body relaxation. 
There are many physical effects that may complicate the model: resonant relaxation \citep{RauchTremaine1996,HopmanAlexander2006,Merritt2015,BarOrAlexander2016}, massive perturbers \citep{Perets2007}, non-spherical geometry \citep{MagorrianTremaine1999,MerrittPoon2004,HolleySigurdsson2006,VasilievMerritt2013,Vasiliev2014}, anisotropy in the initial conditions \citep[hereafter Paper I]{MerrittWang2005,LezhninVasiliev2015}\defcitealias{LezhninVasiliev2015}{Paper I}. All but the last one generally tend to increase the TDE rate compared to the `reference' spherical steady-state value (SSS), and thus would only increase the tension between theory and observations reported in \citetalias{StoneMetzger2016}. The last factor, however, may act in the opposite sense, if the initial distribution of stars was tangentially-biased, having a deficit of stars with low angular momentum. Such a gap could have arised if the galaxy previously contained a binary SBH, which ejected stars from the galactic core through the slingshot process on the way to its merger \citep{MilosMerritt2001}.

In \citetalias{LezhninVasiliev2015}, we explored the effect of gap in the angular momentum distribution on the present-day TDE rates, solving the time-dependent Fokker--Planck equation in spherical geometry. We found that the gap is refilled after $\sim 10^{-2}\,T_\mathrm{rel}$, where $T_\mathrm{rel}$ is the local relaxation time measured at the radius of influence $r_\mathrm{infl}$ (the latter defined as the radius enclosing the mass of stars equal to twice the SBH mass $M_\bullet$). We used \citet{Dehnen1993} models with various values of power-law index $\gamma$ of the density profile and scaled them to real galaxies using the $M_\bullet-\sigma$ relation \citep{FerrareseMerritt2000,Gebhardt2000}, obtaining a one-parameter family of models for each $\gamma$. Using this scaling, the gap-refill time is longer than the Hubble time for $M_\bullet \gtrsim 10^7\,M_\odot$, thus for these galactic nuclei we might expect a reduction of TDE rates compared to the SSS value.

However, the same galaxy merger that creates a binary SBH necessary for the formation of the gap, also leads to a significantly non-spherical shape of the merger remnant. Thus it is important to consider both the shape and velocity anisotropy together in order to obtain a more reliable estimate of TDE rates. The Fokker--Planck method is unsuitable for this task, as the existing implementations are restricted to spherical or at most axisymmetric geometry under certain simplifying assumptions: either that the distibution function depends only on the two classical integrals of motion (the energy $E$ and the conserved component of angular momentum $J_z$), as in \citet{Goodman1983,FiestasSpurzem2010}, or additionally on a third integral that exists only within the sphere of influence, as in \citet{VasilievMerritt2013}. On the other hand, relaxation processes around a SBH can also be studied with the Monte Carlo method, which has been used in spherical geometry by \citet{DuncanShapiro1983,FreitagBenz2002}, but can be easily extended to non-spherical systems \citep{Vasiliev2014}.

The paper is organized as follows. In Section~\ref{sec:theory} we review the basics of the loss-cone theory and the methods used to compute the TDE rates, and in Section~\ref{sec:ic} we describe the initial conditions for our models. In Section~\ref{sec:fp_sph} we apply the spherical Fokker--Planck formalism of \citet{LezhninVasiliev2015} to a sample of galaxies from \citet{Lauer2007}, estimating the TDE rates in the presence of the angular momentum gap for a diverse collection of density profiles. Section~\ref{sec:mc} presents the main suite of our Monte Carlo simulations: we demonstrate that they agree with the Fokker--Planck approach in the case of spherical galaxies, and then consider non-spherical models with and without a gap in the initial distribution. We find that the enhancement of TDE rate due to non-sphericity outweighs the suppression due to the gap, which itself is much less pronounced in non-spherical cases. Section~\ref{sec:results} summarizes our results.

\section{The loss-cone theory and methods}  \label{sec:theory}

The SBH residing at the center of the galaxy directly captures or tidally disrupts stars for which the angular momentum $J$ at pericenter is smaller than the critical value (the loss-cone boundary)
\begin{subequations}  \label{eq:rlc}
\begin{align}
J_\mathrm{LC} &\equiv \sqrt{2GM_\bullet\, r_\mathrm{LC}}\;,\\
r_\mathrm{LC} &\equiv \mathrm{max}\left[ \frac{8GM_\bullet}{c^2}, \left(\eta^2\frac{M_\bullet}{M_\star}\right)^{1/3}\,\!\!R_\star \right] ,\;\;\eta\sim 1,
\end{align}
\end{subequations}
where $M_\star,R_\star$ are the mass and the radius of the star. In the above expression, $r_\mathrm{LC}$ corresponds to either the radius of a direct capture or a tidal disruption; the latter occurs for the main-sequence stars if $M_\bullet \lesssim 10^7\,M_\odot$.
In common with most other studies, we restrict our analysis to the case of a single-mass population of main-sequences solar-mass stars, both as the source of TDEs and as the background providing the two-body relaxation. \citetalias{StoneMetzger2016} have shown that a more realistic mass spectrum moderately enhances the TDE rate. 
 
In a perfectly spherical system, the only process that can change the angular momentum of a star is two-body relaxation, possibly enhanced by massive perturbers and coherent torques (resonant relaxation). In the simplest possible case, a nearly steady-state solution is established after a small fraction of energy relaxation time; the distribution of stars in angular momentum has a nearly logarithmic profile, and thus, owing to the small size of the loss cone, is close to isotropic.
We refer to the review by \citet{Merritt2013} for a more detailed description of the loss-cone theory.
In a non-spherical system, motion of stars is more complicated and often chaotic, at least outside the radius of influence \citep[see, e.g.,][Chapter~4]{MerrittBook}. Most important for the loss-cone problem are the torques that change the angular momentum even in a perfectly collisionless system. In general this leads to a higher TDE rate, especially for less dense galactic nuclei with more massive black holes, for which the collisional two-body relaxation is extremely inefficient.

The evolution of stellar distribution around a SBH has been studied with various methods. $N$-body simulations are the most direct way of incorporating all relevant physical effects, but at present it is impossible to conduct them with realistically large $N$ and realistically small size of the loss cone. Studies such as \citet{Brockamp2011,Zhong2014,Zhong2015} performed simulations with varying number of particles and loss-cone radius, and used empirically derived trends to extrapolate the results to the physically relevant range of values. However, due to complicated boundary conditions (empty vs.\ full loss-cone regimes) the scaling laws even in the SSS case are not trivial and cannot be approximated by ad hoc power-law trends. Moreover, a correct scaling in non-spherical systems must preserve the relative contribution of collisional and collisionless processes to the total relaxation rate; as shown in \citet{Vasiliev2014}, it is mathematically possible, but still places unrealistic demands on the computational cost. 

Next comes the Monte Carlo method for stellar dynamics, which also uses a particle-based representation of the system, but models the relaxation explicitly as a perturbation term added to the equations of motion of particles in a self-consistently generated potential. The code \textsc{Raga} \citep{Vasiliev2015} can deal with moderately non-spherical systems by expanding the potential in spherical harmonics, with coefficients of expansion updated after regular intervals of time, much shorter than the relaxation time, but possibly longer than the dynamical time. This temporal smoothing, together with spatial smoothing indirectly mediated by the functional expansion of the potential, and oversampling (the use of entire particle trajectories to compute the potential, as opposed to positions at a single moment of time), reduces the intrinsic numerical relaxation rate far below the level of conventional $N$-body simulations with a comparable number of particles ($\sim 10^6$). Then a desired amount of two-body relaxation is added to the system; the amplitude of the relaxation term is determined by the number of stars in the physical system being modelled, not by the number of particles in the simulation (the former is typically much larger). The capture radius also can be set to a physically correct value.
These advantages make the Monte Carlo method a very attractive choice for the loss-cone problem. In \citet{Vasiliev2014} it has been applied to more massive SBH ($M_\bullet = 10^7..10^9\,M_\odot$) in galactic nuclei with isotropic initial distribution in angular momentum, neglecting the changes in the density profile; the present study extends it to lower $M_\bullet$ and anisotropic initial conditions, while following the evolution of density profile self-consistently.

Finally, the evolution of stellar distribution function due to two-body relaxation can be described by the orbit-averaged Fokker--Planck equation. As noted above, it is usually applied for spherically-symmetric systems, in which case the distribution function depends on energy $E$, angular momentum $J$ and time. Since the relaxation times are longer than the Hubble time in all but the densest galactic nuclei, we may in the first approximation neglect the diffusion in energy space, and consider only the diffusion in angular momentum separately for each value of energy. This time-dependent PDE has an analytical solution \citep{MilosMerritt2003}, thus allowing a rapid calculation of the flux of stars into the loss cone for any galaxy model (see \citetalias{LezhninVasiliev2015} for more details). Namely, given the density profile of stars $\rho(r)$, we compute the isotropic distribution function $f(E)$ from the Eddington inversion formula, and use it to calculate the energy-dependent drift and diffusion coefficients entering the Fokker--Planck equation. The time-dependent solution of this equation at the given $E$ for any initial conditions $f_0(J)$ is then obtained in the form of truncated series. Finally we integrate over the entire range of energy to obtain the overall TDE rate for the given galaxy. The SSS estimate follows from the same procedure, replacing the time-dependent solution with the expression for the steady-state flux at the given $E$. As the latter is most commonly used in the theoretical context, it is convenient to deal with the ratio $S$ of the time-dependent flux to the SSS value.

\section{Initial conditions}  \label{sec:ic}

Of course, if the time needed to establish a nearly steady-state solution is long compared to the galaxy age, the time-dependent solution will strongly depend on the initial conditions. Here we focus on the case of the `angular-momentum gap', i.e., a deficit of stars with $J < J_\mathrm{gap}$. It is motivated by the following scenario: if the galaxy was formed in a merger of two galaxies, each containing a SBH, then a binary SBH is eventually formed. In order for this binary to reach a small enough separation so that the gravitational-wave emission becomes important, it must shrink its orbit by a factor of $\sim 100$. In the case of a dry merger, the only mechanism that can achieve this is the slingshot interaction between the two SBHs and a third star coming into its vicinity and eventually being ejected, carrying the excess of energy away from the binary. Thus, by the time that the binary coalesces, it inflicts a significant damage to the galactic nucleus, eliminating most of the stars whose angular momenta were lower than the critical value $J_\mathrm{gap}$ -- roughly the angular momentum of the binary itself when it becomes hard. Clearly this value is much larger than the loss-cone size of a single SBH, and this leads to a dramatic suppression of TDE rate until the gap is refilled (\citealt{MerrittWang2005},\citetalias{LezhninVasiliev2015}).

In the present work, we take the initial conditions for galaxies with single SBHs from the end state of simulations of merging binary SBHs, performed in \citet{VasilievAM2015}. In that paper, the same Monte Carlo code \textsc{Raga} was used to follow the evolution of an equal-mass SBH binary with total mass of 1\% of the galaxy mass, from the moment of formation until its coalescence due to gravitational-wave emission, while keeping track of the changing galaxy structure. Three different simulations were conducted under assumptions of spherical, axisymmetric and triaxial geometry; the initial conditions corresponded to a $\gamma=1$ Dehnen profile, and axis ratios were 0.8 in the axisymmetric case, and 0.9 and 0.8 in the triaxial case. As demonstrated in that paper, only in the latter case the slingshot process remains effective enough for the binary to merge in less than a Hubble time even in the absence or with negligible two-body relaxation. Thus for the spherical and axisymmetric systems, our initial conditions are taken from the final snapshot of the simulations where the binary was still far from merger; we imply that some other mechanisms (e.g., gas dynamics) might have driven the binary to coalescence, while it still has had time to carve out the gap in stellar distribution. By contrast, in the triaxial case the end state corresponds to a genuinely merged binary. We also ignore the gravitational-wave recoil that ejects the resulting SBH from the nucleus, and assume that it had enough time to sink back to the center of the galaxy due to dynamical friction \citep[e.g.,][]{GualandrisMerritt2008}.
We stress that these simulations did not follow a merger of two galaxies, but only of two SBHs in a single nucleus; while being somewhat idealized, they allow to explore the influence of geometry in a controlled way. In addition, \citet{VasilievAM2015} also performed a simulation of merging galaxies followed by a formation and merger of a binary SBH, which demonstrated that in a more realistic situation, the evolution of the binary follows the triaxial scenario, even though the deviations from axisymmetric shape of the merger remnant are fairly small. We use the final snapshot of that simulation for a supplementary series of `merger' models.

\begin{figure}
\includegraphics{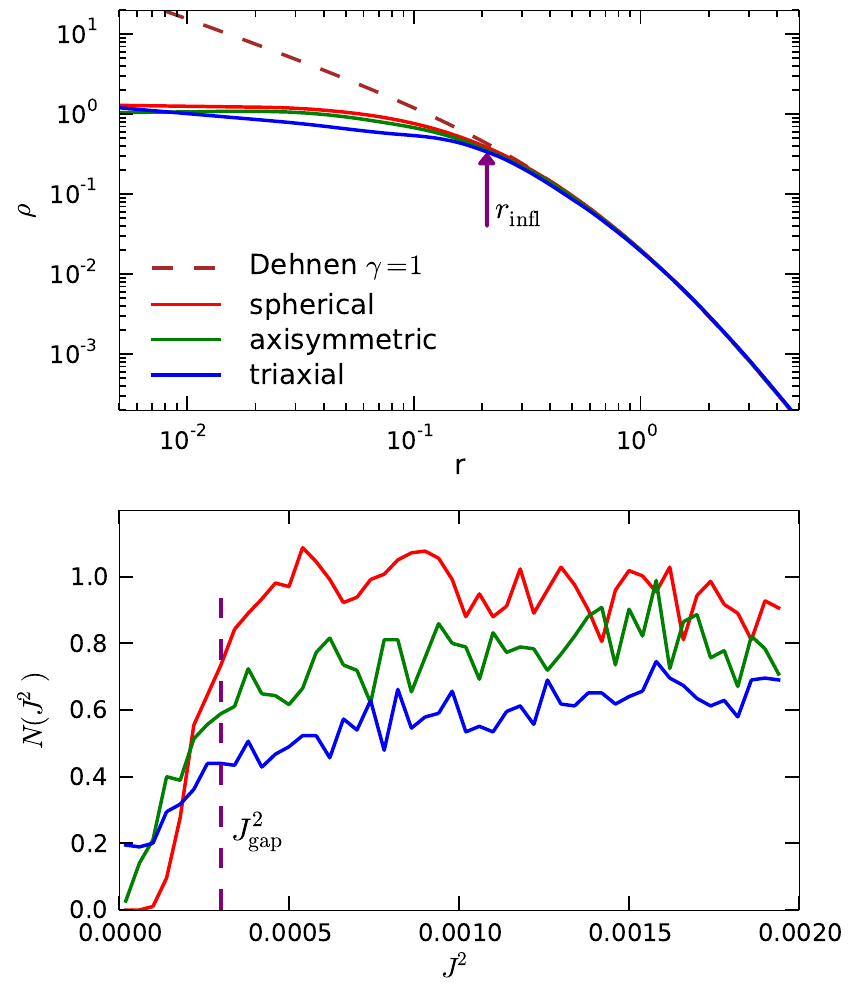}
\caption{Initial conditions for our simulations in three different geometries, taken from the end state of simulations of a binary SBH. Top panel shows the spherically-averaged density profiles, which have almost flat cores inside the influence radius (marked by an arrow), unlike the $\gamma=1$ cusp of the galaxy before the binary-induced evolution took place.
Bottom panel shows the distribution of particles in squared angular momentum, again for three geometries (spherical, axisymmetric and triaxial, from top to bottom). The former one is reproduced from Figure~1 of \citetalias{LezhninVasiliev2015}. The vertical dashed line denotes the boundary of the angular momentum `gap' $J_\mathrm{gap}$; for a spherical system, the distribution function indeed is strongly depleted inside the gap and is almost unperturbed at $J \gtrsim J_\mathrm{gap}$, but for other geometries the depression extends to much larger $J$, but is not as deep inside the gap. At still larger $J$ (beyond the extent of the plot) the three curves converge.
}  \label{fig:ic}
\end{figure}

Our initial conditions are quite different from the original $\gamma=1$ Dehnen models: in addition to the deficit of particles with low $J$, they also have significantly shallower density profiles in the center, with logarithmic slopes close to zero inside the radius of influence (Figure~\ref{fig:ic}, top panel). The distribution function at low $J$ differs significantly between the three cases: while for a spherical system there is a clear gap with almost no particles with $J$ close to zero, in the other two systems the drop is more gradual, but the depression extends to much larger $J$ (Figure~\ref{fig:ic}, bottom panel). The reason is that in non-spherical geometry,
stars interacting with the binary are delivered mostly from centrophilic orbits, in which their time-averaged $J$ may be quite large, but occasionally drops to a low value due to collisionless torques, leading to the slingshot ejection.

The initial snapshots contained $N=0.5\times 10^6$ particles for the three models with predefined geometry, and $N=10^6$ particles for the merger model. For the former ones, we employed a static mass refinement scheme for 2\% of particles with lowest values of angular momentum: each one was replaced with 10 particles of correspondingly smaller mass, bringing their total number to $N=0.6\times 10^6$. Since particles in the Monte Carlo method are moving in the same smooth potential but with independent velocity perturbations, they quickly spread out from their identical initial positions. This refinement allows to improve the statistics of TDE rates, since more numerous smaller particles sample the distribution function more densely and are captured more frequently than in the non-refined case. This is analogous to multi-mass techniques used in collisionless simulations to improve resolution in the region of interest \citep{Zemp2008,ZhangMagorrian2008}; we note that particles of different mass experience the same perturbations, thus heavier particles would not sink to the center due to dynamical friction even after a long time.
We have checked that this refinement scheme does not have any impact on the evolution of the models if loss-cone capture is disabled: both refined and non-refined models are manifestly stable over many thousands of dynamical times, although in the presence of relaxation they slowly develop a \citet{BahcallWolf1976}-type cusp over a timescale comparable to the relaxation time.
A more elaborate scheme of dynamic mass refinement was used in \citet{ShapiroMarchant1978}, where particles were cloned once they cross specific boundaries in the phase space; however we find that in our application a static refinement is sufficient, as most of the captured particles are indeed among the refined ones.

To quantify the effect of the angular momentum gap separately from the overall depression in density profile, we created `isotropized' models with almost the same density profiles%
\footnote{Since it is not possible to have an isotropic model with a density profile shallower than $r^{-0.5}$ if the potential is dominated by the SBH, these models don't have exactly the same profile at radii $\lesssim 0.3r_\mathrm{infl}$.}
as shown in Figure~\ref{fig:ic} and the same shape, but with a nearly-uniform distribution function in $J^2$ (only for models with prescribed geometry). These models were constructed with the Schwarzschild orbit-superposition method, implemented in the \textsc{Smile} code \citep{Vasiliev2013}. The density profiles (spherical or non-spherical) and associated potentials were computed directly from the original $N$-body snapshots and represented in a non-parametric way by a spherical-harmonic expansion with coefficients being arbitrary smooth functions in radius. The same density profiles were then reconstructed with the orbit-superposition method, while placing additional kinematic constraints to enforce uniform distribution in $J^2$. In the rest of the paper, we compare the TDE rates from the models with gap and the isotropized models, not the initial $\gamma=1$ Dehnen models.

\section{TDE rates for spherical galaxies}  \label{sec:fp_sph}

Unfortunately, of the handful of observed TDEs, none occurred in a galaxy where the SBH mass would be known from independent measurements (e.g., equilibrium stellar-dynamical models). Therefore, in assessing the overall rate of TDE, one usually relies on correlations between $M_\bullet$ and other galaxy properties -- velocity dispersion $\sigma$, optical luminocity, etc. 

In \citetalias{LezhninVasiliev2015} we used families of \citet{Dehnen1993} double-power-law density profiles for several values of $\gamma$ with a constant ratio of $M_\bullet$ to the galaxy mass, and all physical quantities scaled according to a particular version of the $M_\bullet-\sigma$ relation. \citet{WangMerritt2004} and \citetalias{StoneMetzger2016}, on the other hand, constructed models based on spherically-symmetric deprojected profiles of individual galaxies, thus their models have a larger variety of profiles and SBH masses. In this section we repeat the analysis of \citetalias{LezhninVasiliev2015} for the same sample of galaxies as used by \citetalias{StoneMetzger2016}, which in turn is mostly taken from \citet{Lauer2007}.

As a first step, we computed the SSS fluxes $\mathcal{F}_\mathrm{SSS}$ following the procedure outlined in the previous section -- deprojection, Eddington inversion, computation of diffusion coefficients, and compared them to those reported by \citetalias{StoneMetzger2016}. As noted in the latter paper, it is not possible to construct non-negative isotropic distribution function for galaxies with too shallow density slopes, thus we discarded such cases, leaving a little over a hundred galaxies in our final sample. The results agreed well with those in \citetalias{StoneMetzger2016}, which is not surprising given that they employed the same workflow, but is encouraging since the implementations are entirely independent.

\begin{figure}
\includegraphics{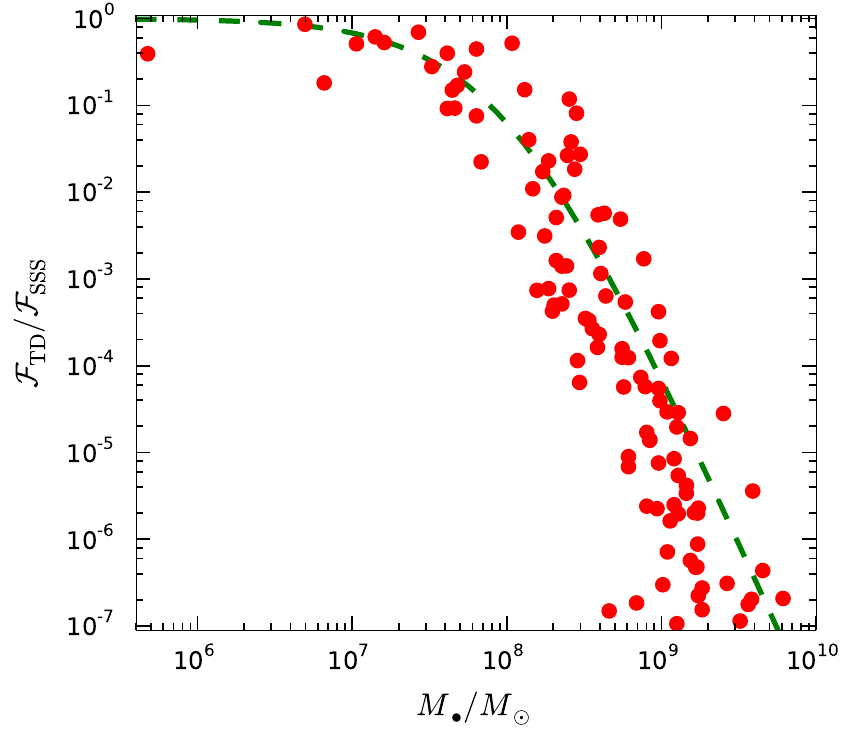}
\caption{Suppression factor $S$ as a function of the SBH mass $M_\bullet$. $S$ is defined as the ratio of TDE rate in the time-dependent solution of the Fokker--Planck equation for a spherical galaxy with a gap in initial conditions to the same quantity computed for the steady-state solution. Points represent the sample of \citetalias{StoneMetzger2016}, in which galaxies have a variety of masses and density profiles; $M_\bullet$ is assigned in that paper from a scaling relation. The dashed line shows the trend $S \sim (1 + M_\bullet/10^8\,M_\odot)^{-4}$, demonstrating that the suppression becomes important for $M_\bullet \gtrsim 10^{7.5}..10^{8}\,M_\odot$.
}  \label{fig:lauer}
\end{figure}

Next we compute the present-day TDE rates $\mathcal{F}_\mathrm{TD}$ from the time-dependent solution of the Fokker--Planck equation, assuming the galaxy age of $10^{10}$~yr and the initial conditions with the gap described by Equation~6 in \citetalias{LezhninVasiliev2015}. Figure~\ref{fig:lauer} plots the suppression factor $S \equiv \mathcal{F}_\mathrm{TD} / \mathcal{F}_\mathrm{SSS}$ as a function of $M_\bullet$. We remind that the SBH mass itself is not measured from observations, but a quantity derived by \citetalias{StoneMetzger2016} from the galaxy properties assuming a somewhat different form of $M_\bullet-\sigma$ relation. It confirms the result of \citetalias{LezhninVasiliev2015}, that in galaxies with $M_\bullet \gtrsim 10^{7.5}..10^{8}\,M_\odot$ the suppression of TDE rate resulting from the gap in angular momentum is significant. However, for the entire cosmic population of SBH, presumably dominated by lower masses, the overall suppression is not very prominent.
\vspace*{10mm}

\section{Monte Carlo simulations}  \label{sec:mc}

\begin{figure}
\includegraphics{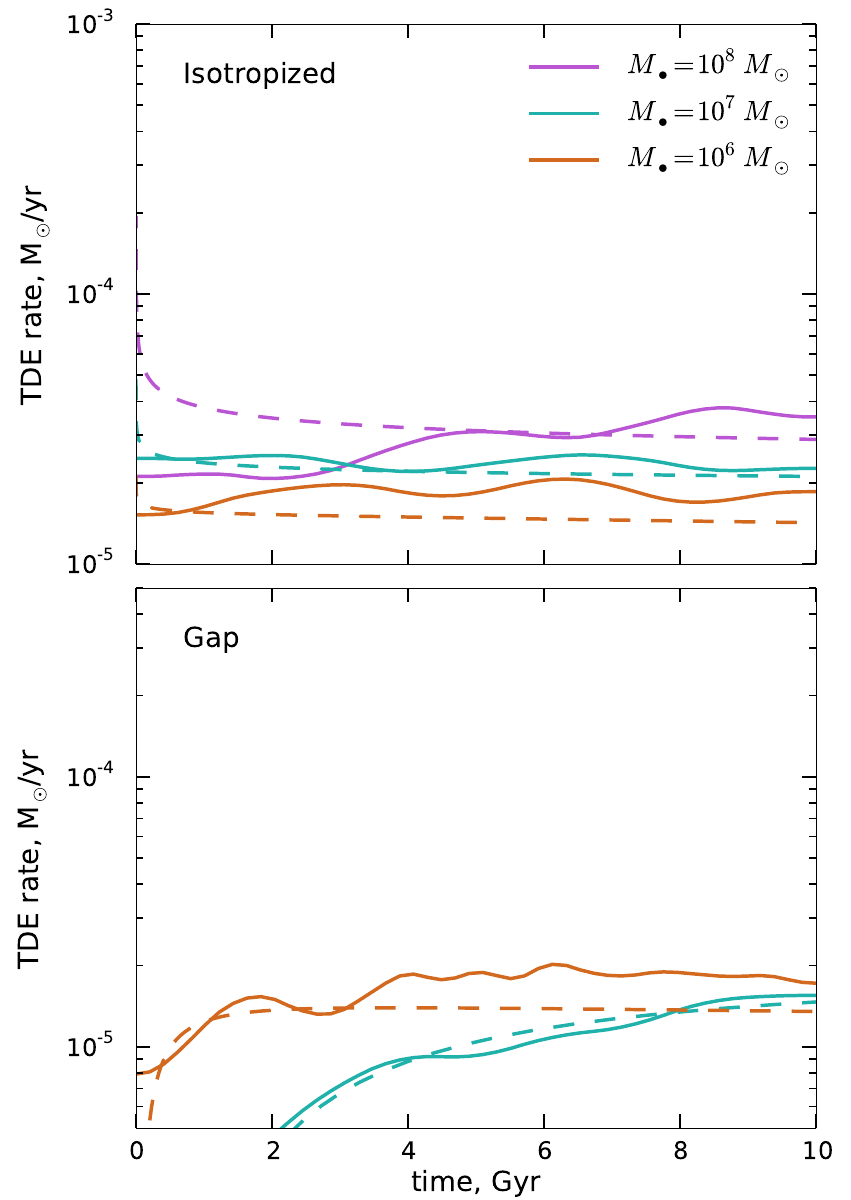}
\caption{TDE rate as a function of time for spherical models with isotropized initial conditions (top panel) and with a gap at low angular momentum (bottom panel). Solid lines are from Monte Carlo simulations and dashed lines are from Fokker--Planck models. Colors denote models with different SBH masses ($M_\bullet=10^8,10^7,10^6\,M_\odot$), in the bottom panel the first one is off scale ($\ll 10^{-6}\,M_\odot$/yr).
}  \label{fig:sph}
\end{figure}

\begin{figure}
\includegraphics{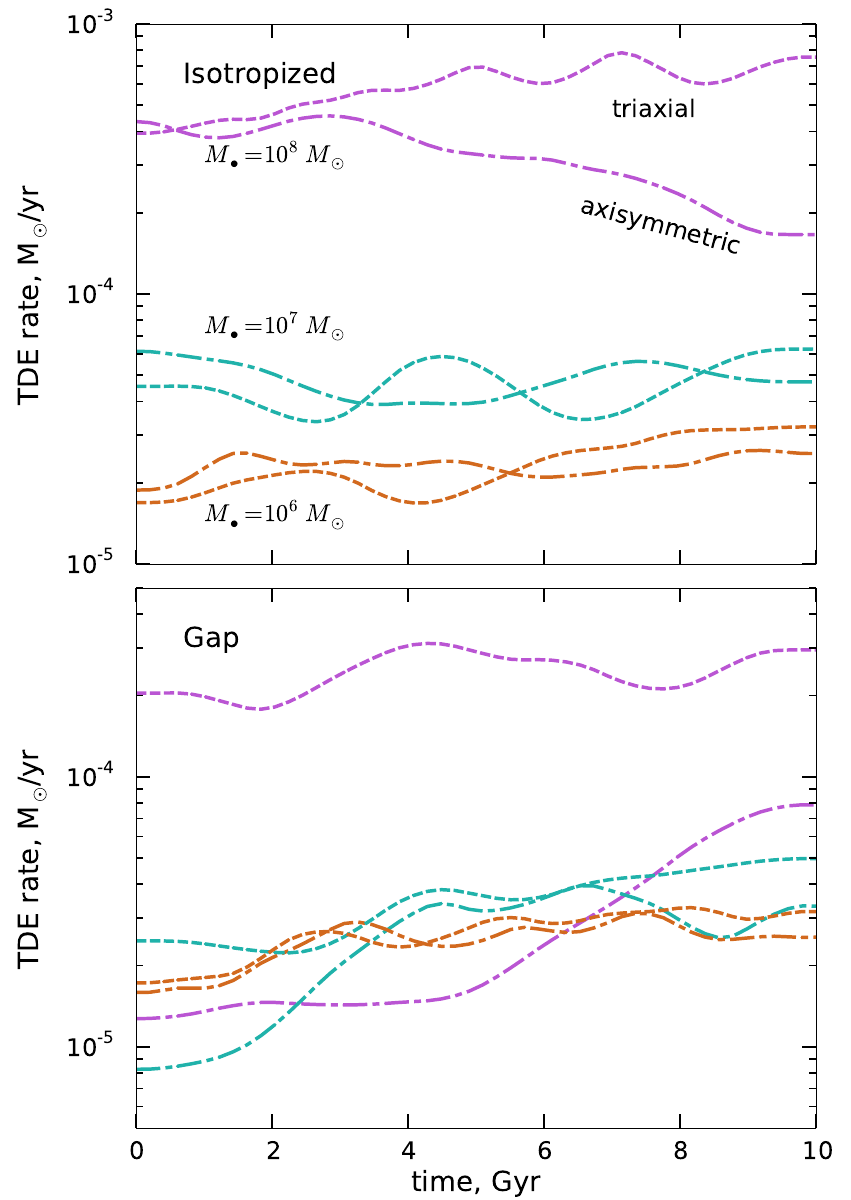}
\caption{TDE rate as a function of time for axisymmetric (long dash-dotted lines) and triaxial (short dashed lines) Monte Carlo models with isotropized initial conditions (top panel) and with a gap at low angular momentum (bottom panel). Colors denote models with different SBH masses (same as in Figure~\ref{fig:sph}). Except for the case $M_\bullet=10^8\,M_\odot$, there is little difference between axisymmetric and triaxial models, since the flux of stars into the SBH is mostly determined by relaxation; however, for the heaviest SBH mass, draining of centrophilic orbits is the dominant mechanism, and there are many more of them in a triaxial system.
}  \label{fig:nonsph}
\end{figure}

Since the Monte Carlo simulations are much costlier than the Fokker--Planck models (but still enormously faster than direct $N$-body simulations with comparable $N$), we cannot afford exploring a hundred of models as in the previous section.
Instead we use the three series of galaxy models with different geometry and initial conditions described in Section~\ref{sec:ic}, scaled to physical units in the following way, similarly to \citetalias{LezhninVasiliev2015}. The SBH mass is assigned a value between $10^6$ and $10^8\,M_\odot$, and the radial scale is determined from the requirement that the radius of influence is
\begin{align}  \label{eq:rinfl}
r_\mathrm{infl} = r_0\;[M_\bullet / 10^8\,M_\odot]^{0.56}\,,\;\;\;r_0 = 35\mbox{ pc}.
\end{align}
The radius of capture or tidal disruption for a star with solar mass and radius (Equation~\ref{eq:rlc}) is $\{2.1, 4.5, 38.4\}\times 10^{-6}$~pc for $M_\bullet=10^6, 10^7, 10^8~M_\odot$, correspondingly, and these are the values that we adopted in our simulations.

We first consider spherical systems, and compare the evolution of TDE rate as a function of time between models with different $M_\bullet$, started from isotropic or depleted initial conditions. In this case we use both the Fokker--Planck method of \citetalias{LezhninVasiliev2015} and the Monte Carlo code \textsc{Raga}; while both share the same prescription for two-body relaxation, they also differ in several aspects. Most importantly, we allow for a self-consistent evolution of the galaxy structure in the Monte Carlo approach, i.e., the gravitational potential changes with time, reflecting the changes in the distribution function. By contrast, in the Fokker--Planck approach we neglect the diffusion in energy and changes in the density profile, which may be noticeable for dense galactic nuclei with short enough relaxation times. Moreover, the mass of stars captured by the SBH is added to $M_\bullet$ in the Monte Carlo simulations, although it always remains small compared to its initial value.
We note that in all Monte Carlo simulations we used the same initial set of models with $N=0.6\times10^6$ particles, only varying the amplitude of two-body relaxation in proportion to $N_\star^{-1} \ln\Lambda$, where $N_\star \equiv 10^2 M_\bullet/M_\odot$ is the number of stars in the galaxy, and $\ln\Lambda\equiv \ln[M_\bullet/M_\odot]$ is the Coulomb logarithm; this is possible because in this approach the relaxation rate can be set independently from the number of particles.

Figure~\ref{fig:sph} shows the TDE rate as a function of time, for isotropized models (top panel) and models with a gap (bottom panel). In the Monte Carlo runs, it is determined by fitting a smoothing spline to the cumulative mass of captured particles as a function of time, and then differentiating this spline, to reduce discreteness noise.
The agreement between Fokker--Planck and Monte Carlo models is remarkably good, although not perfect. For $M_\bullet=10^6\,M_\odot$, the diffusion in energy plays a significant role over Hubble time, leading to a re-growth of a cuspy profile with the logarithmic slope $\gamma \sim 1$, although not as steep as the fully relaxed \citet{BahcallWolf1976} solution. This increase in central density explains higher TDE rates at late times in Monte Carlo simulations, compared to Fokker--Planck models. For $M_\bullet \ge 10^7\,M_\odot$ the changes in density profile can be ignored (we tested this by running simulations in the fixed potential of initial density profile, and found similar amount of captured mass), and the main source of discrepancy are discreteness fluctuations. 

Comparing the spherical models with and without gap, it is clear that for $M_\bullet \ge 10^7\,M_\odot$ the suppression becomes important -- to the extent that in the Monte Carlo simulation with $M_\bullet = 10^8\,M_\odot$ not a single particle was captured over the entire run.

For non-spherical systems we expect the TDE rate to be higher than in the spherical case. First of all, the initial distribution does not have such a strong depression at very small $J$ for these models, as shown in Figure~\ref{fig:ic} (bottom panel). Second, even though the two-body relaxation rate is roughly the same regardless of geometry, the flux of stars into the SBH is higher in nonspherical systems for two reasons. One is that there exist a population of centrophilic orbits, which serve as the `extended loss cone': once a star is on such an orbit even with $J \gg J_\mathrm{LC}$, it will eventually reach the loss cone proper due to collisionless torques, although it may not necessarily pass through pericenter at the moment when $J$ becomes less than $J_\mathrm{LC}$ (this is analogous to the `full-loss-cone' regime of spherical systems, see the discussion in section 4.2 of \citealt{VasilievMerritt2013} for the axisymmetric case). Thus the task of two-body relaxation is to deliver a star into this extended loss cone, and the collisionless torques will do the rest. The other reason is that even without relaxation, the initial number of stars on these centrophilic orbits is quite substantial, especially in the triaxial case, to keep a high `draining rate' for a long time -- possibly longer than the Hubble time. This draining is the dominant contribution to TDE rates in triaxial systems with $M_\bullet \gtrsim 10^8\,M_\odot$ \citep{Vasiliev2014}. For the models with a gap, this population of centrophilic orbits is substantially reduced but not entirely eliminated.

\begin{figure}
\includegraphics[width=8cm]{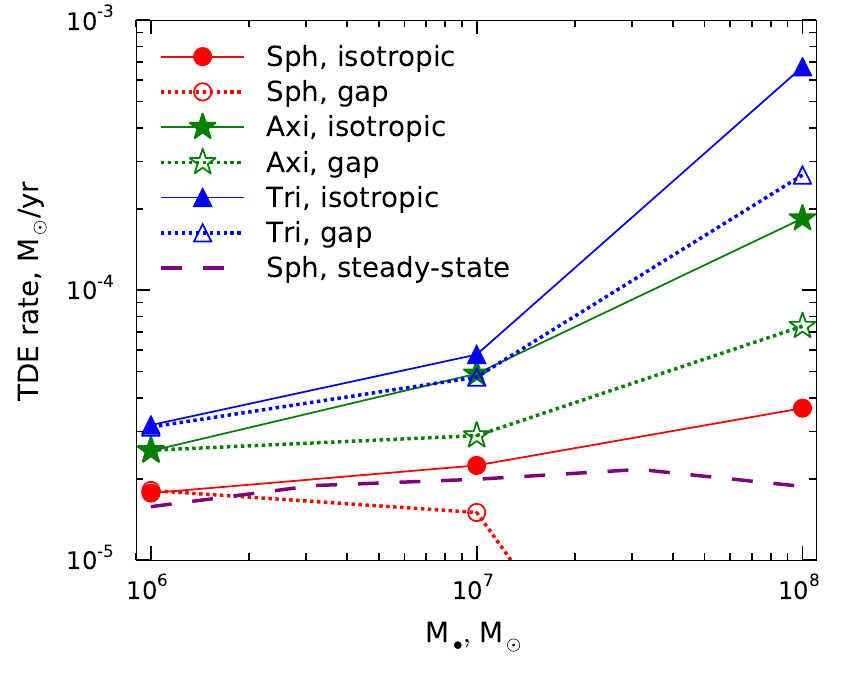}
\caption{Present-day TDE rate as a function of $M_\bullet$ for models with various geometry and isotropized or gap-depleted initial conditions. Red circles -- spherical, green stars -- axisymmetric, blue triangles -- triaxial models; filled symbols -- isotropized initial conditions, open symbols -- models with a gap. Purple dashed line shows the steady-state TDE rate from the Fokker--Planck equation.
The difference between the models becomes more pronounced with the increase of SBH mass. }  \label{fig:summary}
\end{figure}

Figure~\ref{fig:nonsph} shows the TDE rates for axisymmetric and triaxial Monte Carlo simulations, again with isotropized initial conditions (top panel) and with a gap (bottom panel), and Figure~\ref{fig:summary} summarizes the present-day TDE rates (averaged over last 2 out of 10 Gyr of evolution) for all our models. 
The results confirm the above picture: the difference between all three geometries becomes more pronounced with the increase of SBH mass (and hence the relaxation time), because the draining of centrophilic orbits becomes more and more important. Moreover, the suppression of the flux in the models with a gap is nowhere as severe as in the spherical case, and only clearly seen at early times in axisymmetric models with $M_\bullet \gtrsim 10^7\,M_\odot$.
Merger models with initial condition taken from an actual merger simulation were found to be very similar to triaxial models with a gap; we opted not to plot them to avoid crowding.

\section{Discussion and conclusions}  \label{sec:results}

We considered the rate of disruption of stars by massive black holes in galactic centers, with a particular focus on the influence of anisotropic initial conditions resulting from a previously existing binary SBH, and on the role of geometry of the galactic nucleus (spherical, axisymmetric or triaxial). In doing so, we generalized the results of \citet{Vasiliev2014}, where different geometries were considered under the assumption of isotropic initial conditions, and of \citet{LezhninVasiliev2015}, where the effect of a gap at low angular momentum in the initial distribution of stars was studied in spherical galactic models.

Our main tool for this study is the Monte Carlo method for dynamical evolution of stellar systems with arbitrary geometry, introduced in \citet{Vasiliev2015}.  
The initial conditions for the Monte Carlo simulations are taken from models of galactic nuclei with binary SBHs, in which the slingshot process ejects stars with low angular momenta. In order to isolate the effect of angular-momentum gap from the role of geometry, we constructed isotropized models with the same density profile and shape, but having a uniform distribution of stars in angular momentum.

The results can be summarized as follows.
\begin{itemize}[leftmargin=8mm]  \setlength\itemsep{0pt}
    \item For spherical systems, we compared the evolution of TDE rates computed by Fokker--Planck and Monte Carlo methods, and found a satisfactory agreement, confirming our earlier results obtained with the former approach.
    \item Using the Fokker--Planck approach, we computed the expected TDE rates for a sample of galaxies from \citet{Lauer2007} in the presence of a gap in the initial distribution in angular momentum, and compared them with the steady-state estimates from \citetalias{StoneMetzger2016}. We find that the suppression of TDE rates is important for galaxies with $M_\bullet \gtrsim 10^{7.5}\,M_\odot$, again confirming our earlier findings with a more diverse collection of galaxy models.
    \item The depletion of stars with low angular momentum due to slingshot ejection during the binary SBH evolution is less pronounced in non-spherical systems.
    \item The difference between models with isotropic and anisotropic initial conditions and with different geometry starts to play role for $M_\bullet \gtrsim 10^{7.5}\,M_\odot$. Moreover, for non-spherical models, the influence of a gap is less dramatic than for spherical models, and in fact the TDE rate stays well above the SSS estimate (\mbox{$\sim 2\times 10^{-5}\,M_\odot$/yr} under our scaling) for all of them, regardless of the SBH mass and the initial conditions.
    \item Thus we conclude that the effect of tangential anisotropy could be only a minor factor in resolving the discrepancy between theoretical predictions and observationally inferred TDE rates, raised by \citetalias{StoneMetzger2016}, since the majority of TDEs are expected to occur in galaxies with lower SBH masses than our threshold value. The explanation must therefore lie elsewhere, for instance, in the properties of optical emission \citep[e.g.,][]{MetzgerStone2016}.
\end{itemize}

A number of caveats should be mentioned. Our models were scaled to physical units using Equation~\ref{eq:rinfl}, i.e., all properties are fixed by the SBH mass, while for real galaxies this is certainly not the case. Moreover, in the range of $M_\bullet$ considered in this paper, SSS estimates of TDE rates for the sample of galaxies in \citetalias{StoneMetzger2016} are up to an order of magnitude higher than our values $\sim 2\times10^{-5}\,M_\odot$/yr, which indicates that our adopted scaling (taken from a series of previous papers) underestimates the density of real galaxies in the low-mass end. Thus the TDE rates shown in Figure~\ref{fig:summary} should be interpreted as lower boundaries on the range of values possible in real galaxies, and a typical galaxy would have a TDE rate a factor of few higher. As illustrated by \citet{StoneVelzen2016}, some E+A galaxies may have TDE rates still an order of magnitude higher than average, owing to their high central density. This also means that the effects of non-sphericity and angular-momentum gap would be noticeable starting from a somewhat higher $M_\bullet$ than in our simulations, since a denser galaxy has a shorter relaxation time and thus is less affected by these factors.

Our initial conditions are taken from simulations of merging binary SBHs conducted in \citet{VasilievAM2015}. The main suite of simulations were set up not in a cosmological context, but rather embedded an equal-mass binary SBH into a steady-state galaxy with a prescribed shape. This is of course not very realistic, but allows a more precise control on the role of geometry in both the evolution of the binary SBH, and the subsequent loss-cone dynamics around a single SBH. The models considered here and in that paper are only moderately non-spherical, with minor-to-major axis ratio $\gtrsim 0.75$ and intermediate-to-major axis $\gtrsim 0.9$.
We also considered a series of `merger' models using initial conditions from a simulation that followed a merger of two galaxies followed by formation and merger of SBHs. The shape of this model was close to but not precisely axisymmetric. Nevertheless, even this small deviation brings qualitative changes to the structure of orbits around the SBH, and the TDE rates for these models were very similar to the triaxial ones.
Our models also have a very shallow density profile in the center, resulting from the destruction of the pre-existing cusp during the merger. Galaxies with a steeper density profile ($\gamma \gtrsim 1$) probably either have not experienced a major merger in their history, or have re-grown the cusp due to a short relaxation time or due to dissipative processes (inflow of fresh gas and star formation). In all these cases we don't expect the angular-momentum gap to exist.

To summarize, simple estimates of TDE rates under the assumption of spherical geometry, steady state and isotropic initial conditions, commonly used in the literature, should be regarded as a lower bound for the TDE rates expected in more complex systems resulting from galaxy mergers, because the effect of angular-momentum gap is more than compensated by a non-spherical shape of the merger remnant. We remind that the software for computing the steady-state and time-dependent TDE rates in spherical geometry for an arbitrary density profile, using the Fokker--Planck approach, is available at \url{http://td.lpi.ru/~eugvas/losscone}.

\section{Acknowledgements}  \label{sec:discussion}

EV acknowledges support from the European Research council under the 7th Framework programme  (grant No.\ 321067) and from NASA (grant No.\ NNX13AG92G).
This work was partially supported by Russian Foundation for Basic Research (grant No.\ 15-02-03063).
We thank Nick Stone for providing the data for the calculations performed in \citetalias{StoneMetzger2016}, and for a fruitful discussion.

\label{lastpage}

\end{document}